# A Mobile Message Scheduling and Delivery System using m-Learning framework

Moumita Majumder, Sumit Dhar

**Abstract**— Wireless data communications in form of Short Message Service (SMS) and Wireless Access Protocols (WAP) browsers have gained global popularity, yet, not much has been done to extend the usage of these devices in electronic learning (e-learning) and information sharing. This project explores the extension of e learning into wireless/ handheld (W/H) computing devices with the help of a mobile learning (m-learning) framework. This framework provides the requirements to develop m-learning application that can be used to share academic and administrative information among people within the university campus. A prototype application has been developed to demonstrate the important functionality of the proposed system in simulated environment. This system is supposed to work both in bulk SMS and interactive SMS delivery mode. Here we have combined both Short Message Service (SMS) and Wireless Access Protocols (WAP) browsers' .SMS is used for Short and in time information delivery and WAP is used for detailed information delivery like course content, training material, interactive evolution tests etc. The push model is used for sending personalized multicasting messages to a group of mobile users with a common profile thereby improving the effectiveness and usefulness of the cntent delivered. Again pull mechanism can be applied for sending information as SMS when requested by end user in interactive SMS delivery mode.
The main strength of the system is that, the actual SMS delivery application can be hosted on a mobile device, which can operate even when the device is on move.

——————————— ◆ ———————————

## 1 INTRODUCTION

It is expected that PDA/mobile phone sales will be win over PC sales in India and North East India will also be part of this. Computing devices have become ubiquitous on today's college campuses. From notebook computers to Wireless phones and Handheld devices (or W/H devices for short), the massive infusion of computing devices and rapidly improving Internet capabilities have altered the nature of information sharing within university campus. It combines individualized (or personal) learning with anytime and anywhere delivery of information. With a W/H device, the relationship between the device and its owner becomes one-to-one, always on, always there, location aware, and personalized. M learning application can always reach the audience and vice versa. Any urgent information can be delivered in almost real time manner. On the other hand, scheduling the delivery time of information according to convenience by developing an efficient and intelligent SMS delivery System is part of present work. This work explores the integration of mobile technology for data services like Short Message Service (SMS), Wireless Access Protocols (WAP) and Wireless Markup Language (WML) for information sharing with in the university campus.

————————————————

- *Moumita Majumder is with the Department of Computer Science and Engineering, Tripura University, India*
- *Sumit Dhar is with the Department of Computer Science and Engineering, Tripura, India*

The WAP and SMS gained global popularity for data services because of its thin-client architecture and device independence. The thin-client architecture allows applications to run on the server and transported to W/H devices thereby removing the need for sophisticated client device. Despite W/H device popularity with students, not much has been done yet to extend e learning to these devices in India. So, present work simulates an environment where university administration can be controlled by using mobile technology.

## Mobile learning technologies:-

Among all the technologies SMS is the simplest of all technologies, and interactive learning activities can be devised with very basic equipment.
As specified by Mellow SMS technology may be employed according to three models within m-learning [1]. These will be in the form of:
- A 'push' system, where the institution pushes out messages to all students in a course.
- A 'pull' system, where students order specific information based on a menu of all listed content on a web page or a paper handout.
- An interactive model, where questions are either sent out or ordered, then answered, and replied to by the student to check the answers and receive feedback.

Mellow also identified key advantages of SMS to students as follows [1]

- pace of their learning.
- Specificity of content.
- Tutor-constructed study aids designed for those areas that are 'challenging to learn'.
- Using technology that is engaging and totally comfortable for the student.
- Non-threatening, private availability of on-demand study support.

Krassie Petrova described an experiment involving mlearning using SMS, in his paper "Student revising for a test using SMS"[2]. He also explored that most of the reviewed scenarios are not tied to a particular event timeframe and are driven by the provider ('push' mode) rather than by the learner. But in his discussion he discussed an SMS scenario which is both learner-driven ('pull' mode) and context dependent (student are studying for a pre-scheduled assessed test).

Present worker presents an information sharing approach within the University campus using mobile technology framework. Architecture of the prototype system has been described, developed and simulated.

## 2 RELATED WORK:

Two studies at European universities have focused exclusively on use of SMS technology as collaboration tools for m-learning. The first study emulated a W/H device on a PC to allow students send SMS messages on various discussion topics, which were aggregated and categorized by the instructor, using an electronic whiteboard, in the classroom [3]. The categorization can be done by criteria such as sender, receiver, time and others. The second study evaluated the effectiveness of SMS campaign as a conversational mechanism in context of developing better quality mobile teaching and learning Environment [4]. The effectiveness SMS campaign was measured by quickness of the response, the quality of data collected, the impact of message complexity on number of responses and the method of campaign announcement on quality and quantity of messages. These studies demonstrated that students liked using SMS and they were responsive to the use of W/H devices for interaction and learning. The response rates were high and the quality of the messages was very good. SMS responses were also much quicker than email responses. Both these studies experiment with popular mobile messaging services to see whether they would work in m-learning environment and provide support for the conversational theory of learning [5].

The University of Pretoria in South Africa started using mobile phone support during 2002 in three paper-based distance education programs because more than 99% of the 1,725 students (2002) had mobile phones [6]. The majority of these learners live in remote rural areas with little or no fixed-line telecom infrastructure. Mobile phone support to these rural distance-learning students entails sending bulk, preplanned SMSs to specific groups of students extracted from the database for specific administrative support (customized group SMSs). The advantages and successes have already been significant. In response to a reminder for registration for contact sessions, 58% of the learners registered before the closing date, compared with the normal expected percentage of below 40%. In response to a reminder of the contact session dates, 95% of the learners who had registered for the contact sessions attended. Learners respond in mass and almost immediately to information provided in SMS messages.

Present work differs from the above in the dissemination architecture as it uses J2ME platform and XML messaging. Using same tools another architecture [7] has been designed for delivering rich content into mobile phones. But in that case of Rich Content Mobile Learning, a J2ME smart client application is needed to be installed in every end user device. It is very difficult to distribute and install the client software to each and every end user. The adaptability of the client application to different types of end user devices and operating systems is also an issue.

In the present architecture, the J2ME application is needed to be installed on single mobile phone, which makes the final delivery of the SMS to all end user mobiles. Also in the existing architecture[7], the J2ME smart client needs to be installed on end user device and the device will communicate with the web service directly over GPRS for exchanging information. So, all the end user devices are needed to be J2ME and GPRS enabled. These types of high-end mobile devices are not readily available in less developed part of our country as Tripura. In present architecture, the end users need only basic devices with SMS over GSM facility as they are needed to send and receive SMS only. The actual communication with web service and processing of XML information is done by the single mobile phone on which the J2ME application will be installed. This mobile phone will send back the processed information in the form of SMS to end user devices. Thirdly in the existing architecture [7], the actual transfer of XML content will be done between the end user and the web service. So the end user has to pay for the exchange of high data transfer. But in our architecture, the end user will request and receive information in the form of SMS only. So it will be a less costly affair. The existing system deals with different types of media (text, image, audio, and video). Our system uses the SMS platform. So volume and depth of information will be less in it. But it will be a far more practical solution in the context of availability of end user devices with lesser configurations.

## 3 REASON TO DEVELOP A NEW MOBILE MESSAGE SCHEDULING AND DELIVERY SYSTEM

While studying different SMS delivery systems used in different existing m-learning projects, it has been felt that there is urgent need to develop a system to meet the following University requirements

<u>Administrative learning support</u>

- Access to administrative information.

- Access to examination and test marks via a mobile service number or m-portal.
- Access to financial statements – fees etc.
- Registration data via mobile service number or m-portal.

<u>Academic learning support</u>

- Communication and interaction (bulk short message service (SMS) and interactive voice response).
- Assessment (Multiple choice questions/quizzes).
- Feedback on assignments and tasks from Teachers.
- Motivational and instructional messages.1
- Multicasting of important learning activities like seminars, guest lectures etc.

The above requirements are the main driving force to develop a new SMS scheduling and delivery system.

## 4 DISSEMINATION ARCHITECTURE

### The system Description

### The SMS push method:

The wireless messaging system will generate and send SMS containing information to the concerned recipient. This system can be used to broadcast important university events like exam schedule, consolidated result, detailed schedule of seminar and visiting faculty and many more. There are other long waited requirements, which this system will meet. One of them is to act as means of administrative communication between authority and various department heads like invitation for an important meeting. A new all-round library system is currently being developed and going to be deployed in near future. Our system can integrate with the system and provide library services like new book alert and defaulter alert.

### The SMS pull method:

Besides that, an interactive SMS service is also an important functionality provided by the system. This system can be used to render information on demand like detailed result, course material, reference book etc while operating on interactive SMS mode. The recipient can request for particular information by sending the specific code required in the form of SMS to the system and in reply the information will be delivered again in form of SMS. As an example, to know the consolidated result one is needed to send "<result > space <enrolment No>" to the system.

### SMS Scheduling:

The System is capable of SMS scheduling so that the SMS to be delivered can be scheduled to be sent on any future point of time.

### The Web/WAP site:

A web/ WAP site has been currently under development. It will be used for detailed information delivery like course content, training material, interactive evolution tests etc. One of the main functionality of this system is to provide SMS configuration screen through which the authorized user can create and schedule new SMS. The site will be accessible both over web and WAP platform.

**The detailed system is represented in the below block diagram**:

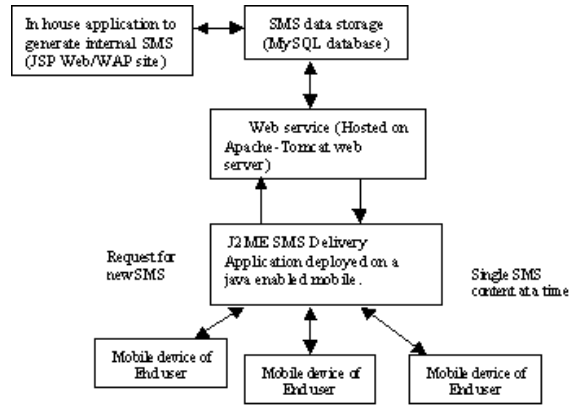

The SMS information will be stored in a central SMS database table. A web/WAP application hosted in university intranet will work as internal SMS feed to the data storage.
One J2ME SMS delivery application will query one web service, which will provide SMS information from the data storage. The delivery application can be hosted on a mobile device.
Here the data storage will be one MySQL database. The final delivery component will be the J2ME application hosted on a mobile, which will actually send the SMS to mobile devices of end users. The web service will work as communication link between these two tiers.

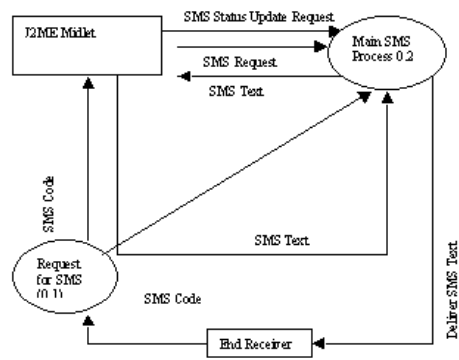

**Data Flow Diagram (Interactive SMS, PUSH and PULL method)**

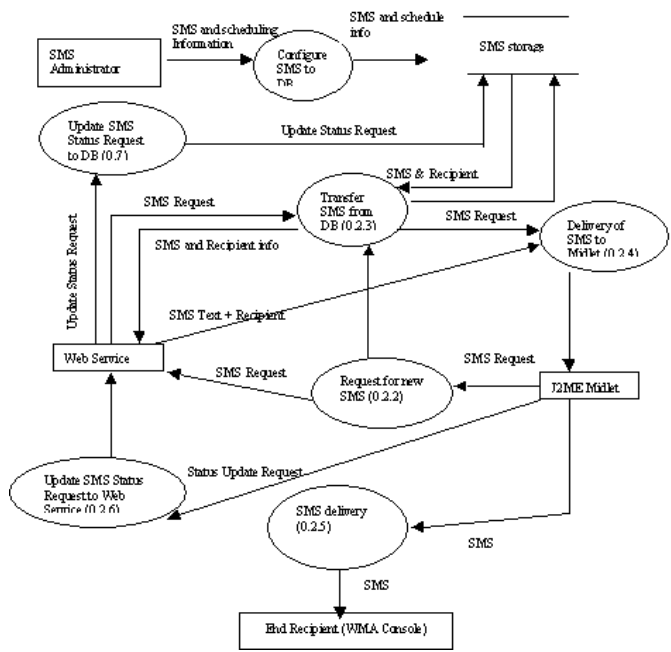

**DFD Push SMS Delivery**

## 5 DESCRIPTION OF THE SYSTEM IN SIMULATE ENVIRONMENT(IMPLEMENTATION)

The three main components of the system are MySQL database named Informatica, one web service named TUSMSDataService and one J2ME application called SMSClient. Informatica DB is hosted in a MySQL database. The TUSMSDataService is hosted in an Apache Tomcat web server. While executing, the SMSClient will be deployed and running in an Emulator provided by Wireless Toolkit 2.5.2. The SMSClient application will continuously communicate with the web service to check the availability of new SMS once invoked by pressing the "SendSMS" button on the Midlet Screen on the Emulator. The TUSMSDataService in terms will communicate with the Informatica DB to fetch SMS information. SMS content will be configured by Administrator into the Informatica DB.

Once new request for new SMS is submitted to TUSMSData-Service by SMSClient, the web service will fetch new SMS information from DB if available and then deliver to SMSClient. SMSClient will then deliver the SMS content to the end recipient. The recipient information is included in the SMS information sent by the Web Service.
Once the SMS is sent, SMSClient will report the success/failure status to the Web Service and the Web Service in terms update the status to Informatica DB. This process will continue till stopped by pressing "Exit" button on the Midlet screen.

The SMS will be received by WMA consoles in the simulated environment

## 6 CONCLUSION

**Economic viability:**

One apparent advantage of this application is that the students did not have to pay anything for the mobile data service, which in some studies has been identified as one hindrance for this type of applications. This facility provided by the service can be treated as a part of student welfare and running expenditure can be arranged from the same fund.

**Anytime and anywhere administration:**

One of the most prominent features of the architecture is that, the actual delivery component (the J2ME application) is hosted on a mobile. As a result, to initiate the SMS sending process, the administrator needs to just switch on the mobile device and run the J2ME application. The application then will communicate with the web service and starts sending the unsent SMSs even when it is on move.

Again, using the same mobile device, the administrator can access the SMS configuration screen of the JSP site and create and schedule new SMS content.

**Scalability:**

The J2ME application can be hosted on as many mobile devices as required. Each of the mobile phone will act as SMS delivery component. They will communicate with the web service individually and perform the bulk-sending task in parallel way. A status flag is associated with each SMS entry in database, which will be useful in concurrent delivery model. There will be 3 status values i.e. 0: New, 1: processing is going on for delivery and 3: Successfully sent. So if one delivery process is processing one SMS it will get the status 1 and will not be processed by other process. Thus efficiency and scalability of the overall system can be increased.